\documentstyle[array,multicol,aps,preprint]{revtex}
\tightenlines

\newcommand{\eref}[1]{(\ref{#1})}

\begin{document}
\preprint{{hep-th/0206124} \hfill {UCVFC-DF-35-2002}}
\title{The space of signed points and the Self Dual Model}
\author{Lorenzo Leal{${}^{a, b, }$}\footnote{lleal@fisica.ciens.ucv.ve}}
\address{${}^a$Grupo de Campos y Part\'{\i}culas, Departamento de F\'{\i}sica, Facultad de Ciencias, Universidad
Central de Venezuela, AP 47270, Caracas 1041-A, Venezuela \footnote{permanent address}\\
${}^b$Departamento de F\'{\i}sica Te\'orica, Universidad
Aut\'onoma de Madrid, Cantoblanco 28049, Madrid, Spain}
%

\maketitle

\begin{abstract}

We study a generalization of the group of loops that is based on
sets of signed points, instead of paths or loops. This geometrical
setting incorporates the kinematical constraints of the Sigma
Model, inasmuch as the group of loops does with the Bianchi
identities of Yang-Mills theories. We employ an Abelian version
of this construction to quantize the Self-Dual Model, which
allows us to relate this theory with that of a massless scalar
field obeying nontrivial boundary conditions.
\end{abstract}

\section{Introduction}
It is well known that there is no natural way of defining
non-Abelian theories of $p-$forms, for $p>1$. This is closely
related to the lack of a simple notion of order for $p-$surfaces
($p>1$). Therefore, besides the path-space representations of
gauge theories \cite{uno,dos}, it only remains, in the non-Abelian
case, the possibility of considering "$0-$surfaces", i.e.,
points, as the underlying objects to enter in a geometric
representation of quantum field theories. In this paper, we shall
study this problem. Despite these ideas are motivated by
considerations about non-Abelian theories, we find it convenient
to present, as an example of their application, the
"signed-points" representation of the Self-Dual Model
(SDM)\cite{tres}, since  it appears that certain properties of
this Abelian model are conveniently displayed in this geometrical
framework. More preciselly, we find that the SDM can be seen as
the theory of a massless scalar field that obeys anyonic boundary
conditions. This agrees with an earlier result about the
Maxwell-Chern-Simons theory (MCST) \cite{cuatro} (which is dual to
the SDM \cite{cinco}), that was obtained by working in a
path-representation \cite{seis}. The latter representation has
also been recently used with the SDM, in order to study the
geometrical content of the duality symmetry between this model
and the Maxwell-Chern-Simons one \cite{seisprima}.

 In the next section we present
the general ideas about the signed-point space. In the last one
we discuss their application to the SDM.

\section{The space of signed points}\label{sec2}

Consider the space  of ordered lists of points in $R^n$ (the
extension to general manifolds is immediate). We shall declare
that there are to kinds of points, that we arbitrarily take as
positive (or "points") and negative ("antipoints"). A typical
element $X$ of this space can be represented by

\begin{equation}
X =
(x_{1}^{(s_1)},x_{2}^{(s_2)},x_{3}^{(s_3)},...,x_{r}^{(s_r)}),\label{uno}
\end{equation}
where the "sign" $s_a= \pm$ of each point $x_a$ has been
explicitly written. The number of points $r$ is arbitrary. We
define the composition $X.Y$ of two lists as

\begin{equation}
X.Y =
(x_{1}^{(s_1)},x_{2}^{(s_2)},...,x_{r}^{(s_r)},y_{1}^{(t_1)},y_{2}^{(t_2)},...,y_{u}^{(t_u)}).
\label{dos}
\end{equation}
The space of lists, with the composition defined above, can be
endowed with a group structure as follows. Firstly, we demand
that pairs "point-antipoint" be annihilated if they meet at the
same place and consecutively in a list. For instance,
$(x_{1}^{(+)},x_{2}^{(+)},x_{2}^{(-)},x_{3}^{(+)})$ will be taken
as $(x_{1}^{(+)},x_{3}^{(+)})$. Once all the consecutive and
equally located pairs in $X$ have been annihilated, we are left
with a "reduced list" (RL) of "signed-points" $R(X)$. The product
of two RLs is then defined as the RL associated to their
composition

\begin{equation}
R(X_1).R(X_2) \equiv R (R(X_1).R(X_2)) .\label{tres}
\end{equation}
It can be seen that that the space of RLs forms a group under the
multiplication  defined by eq. \eref{tres}. The identity element
results to be the empty list. The inverse element $R^{-1}(X)$ is
the reduction of the list built by inverting the order and
changing the signs of the points that appear in $X$:

\begin{eqnarray}
R^{-1}(X)=
R(x_{r}^{(-s_r)},x_{r-1}^{(-s_{r-1})},...,x_{1}^{(-s_1)}).\label{cuatro}
\end{eqnarray}

 Next, we are going to consider functionals $\Psi(R(X))$ that
 depend on RLs. To simplify the notation, we shall label
 $R(X)$ simply as $X$. This should not lead to confusion, since
 from now on we shall restrict ourselves to deal with RLs. We define an operator
 $a(Y)$ that appends a RL $Y$ to the left of the argument $X$ of
 the RL-dependent functional $\Psi(X)$
\begin{equation}
a(Y)\Psi(X) \equiv \Psi(Y^{-1}.X) .    \label{cinco}
\end{equation}
Consistence demands that

\begin{eqnarray}
 a(X_1)a(X_2)\Psi(X)& = & \Psi(X_{2}^{-1}.X_{1}^{-1}.X))\nonumber\\
 & = & a(X_1 .X_2 )\Psi(X) \label{seis},
\end{eqnarray}
hence, $a(X)$ constitute a representation of the group of RLs,
acting on RL-dependent functionals. Furthermore, since

\begin{equation}
a(X) = a(x_{1}^{(s_1)})\,a(x_{2}^{(s_2)})\,...\,a(x_{r}^{(s_r)}),
\label{siete}
\end{equation}
the $a's$ depending on a single point generate the representation.
Observe that

\begin{equation}
  a(x^{(+)})=  a^{-1}(x^{(-)}),\label{ocho}
\end{equation}
thus, we can adopt the notation $a(x)=  a(x^{(+)})$ (and
$a^{-1}(x)= a(x^{(-)})$) without ambiguity. It is worth observing
that besides being RL-dependent operators, the $a's$ are ordinary
functions of the variables $x_a$, $a=1,...,r$, and can be, for
instance, derived with respect to them.

As in the Group of Loops case \cite{uno}, there exists a sort of
infinitesimal generators that we define as follows. Take the RL
$\delta Y$, consisting on a pair point-antipoint, separated by an
infinitesimal vector $u$
\begin{equation}
 \delta Y =((x+u)^{(+)},x^{(-)}).\label{nueve}
\end{equation}
In the limit $|u|\rightarrow 0$, this "dipole-list" reduces to
the identity. In this sense it is an infinitesimal element of the
group of RLs. We define $\delta_{\mu}(x)$ as the operator that
measures the response of $\Psi(X)$ when its argument $X$ is
slightly changed by appending the "dipole" $\delta Y$ at $x$

\begin{equation}
 \Psi(\delta Y .X)- \Psi(X)\equiv u^{\mu}\delta_{\mu}(x)\Psi(X) ,\label{diez}
\end{equation}
up to first order in $u$. From equations \eref{cinco} \eref{ocho}
\eref{nueve} \eref{diez}, one has
\begin{eqnarray}
(1+u^{\mu}\delta_{\mu}(x))\Psi(X) &=&
a(x)\,a^{-1}(x+u)\Psi(X)\nonumber\\
&=& a(x)\Big( a^{-1}(x)+u^{\mu} \frac{\partial}{\partial
x^{\mu}}a^{-1}(x) \Big)\Psi(X)\nonumber\\
&=& \Big( 1+u^{\mu}a(x)\frac{\partial}{\partial x^{\mu}}a^{-1}(x)
\Big) \Psi(X);\label{once}
\end{eqnarray}
thus, we obtain the identity

\begin{equation}
\delta_{\mu}(x)\equiv a(x)\frac{\partial}{\partial
x^{\mu}}a^{-1}(x), \label{doce}
\end{equation}
and we see that the "dipole derivative" $\delta_{\mu}(x)$ can be
analyzed in terms of the elementary generators $a(x)$. From eq.
\eref{doce} it is immediate to obtain

\begin{equation}
\partial_{\mu}\delta_{\nu}(x)- \partial_{\nu}\delta_{\mu}(x)+
[\delta_{\mu}(x),\,\delta_{\nu}(x)]=0 ,\label{trece}
\end{equation}
which is just the kinematical constraint obeyed by chiral fields.
This should be compared with the Bianchi identity obeyed by the
area derivative in the loop-space formulation of
Gambini-Tr\'{\i}as \cite{siete}, \cite{sieteprima}. There is a
geometric construction underlying identity \eref{trece} that
deserves to be pointed out. Take an infinitesimal parallelogram
of sides $u$,$v$, centered at $x$. At each vertex, put a pair
point-antipoint. The resulting configuration is then equal to the
empty list which in turn, is the identity of the group. On the
other hand, the same configuration can also be reached by a
successive pasting of "dipoles": the first one with its point at,
say, $x+u$, and its antipoint at $x$, the second one consisting
on a point at $x+u+v$ and an antipoint at $x+u$, and so on. Since
the two constructions correspond to the same RL, namely, the
identity of the group, one has:

\begin{equation}
(1+u^{\mu}\delta_{\mu}(x))(1+v^{\mu}\delta_{\mu}(x+u))(1-u^{\mu}\delta_{\mu}(x+u+v))(1-v^{\mu}\delta_{\mu}(x+v))=
1, \label{catorce}
\end{equation}
and it is a trivial matter to see that this is the same as
eq.\eref{trece}, up to first order in the area of the
parallelogram expanded by $u$ and $v$.

Summarizing, we see that the RLs or "signed-points" space,
encodes, through its infinitesimal generators, the kinematical
properties of chiral (or sigma-model) fields. This results as a
consequence of the group structure, and follows strictly from
geometrical considerations.

This construction, like the very definition of the space of RLs
and its generators, is very close to the Loop-Space construction
of Gambini-Tr\'{\i}as \cite{siete}, \cite{sieteprima}, which is
the basis for the present formulation.

\section{An application: Self Dual Model and the Abelian Group of Signed Points}\label{sec3}

In this section we present a simple application of the ideas
discussed above. In a recent article that deals with the
quantization of the Maxwell-Chern-Simons  Theory (MCST) in a
geometric representation \cite{tres}, it was mentioned that an
appropriate geometrical setting that would serve to relate the
topological interaction provided by the Chern-Simons term, with
certain anyonic behaviour of the wave functional of the theory,
should be one of "points and antipoints" (in the sense discussed
before), both for the (MCST) and its dual model (which is the
SDM). This conclusion was reached after solving the "Gauss
constraint"
 of the theory in a path-representation, and noticing that the
 feature of the paths that survives in the reduced phase space is precisely the distribution
 of their ending points. These boundary-points acquire a long-range
 interaction due to the topological term.  Now we address this point in some detail, providing an example of
how the ideas of the preceding section could be useful in Field
Theory. We shall restrict ourselves to the SDM, since it is in
this model where the RLs representation can be implemented in a
more natural and geometrically appealing form.

We start from the SD action in the Stueckelberg form

\begin{equation}
S = \int d^{3}x \,\Big ( \frac k2\,\,
\varepsilon^{\alpha\beta\gamma}\partial_{\alpha}A_{\beta}A_{\gamma}
+ \frac 12
(A_{\alpha}+\partial_{\alpha}\varphi)(A^{\alpha}+\partial^{\alpha}\varphi)\Big
) ,\label{quince}
\end{equation}
which is invariant under the gauge transformations
\begin{equation}
 A_{\alpha}\rightarrow A_{\alpha} +\partial _{\alpha}\Lambda,\label{dieciseis}
\end{equation}

\begin{equation}
 \varphi \rightarrow \varphi - \Lambda .\label{diecisiete}
\end{equation}
The equations of motion that follow from varying $S$ w.r.t.
$\varphi$ are nothing but consistence equations for the true
equations of motion, that result when varying w.r.t. $A_{\alpha}$.
This reflects the unphysical character of the Stueckelberg field
$\varphi$, which could be set equal to zero by a gauge choice,
accordingly with eq. \eref{diecisiete}. Instead, we are interested
in "gauging away" the Chern-Simons field, in a sense that will be
clear soon, and within the spirit of what is a common procedure
in ordinary quantum mechanics of particles with Chern-Simons
interactions \cite{ocho}.

The quantization in the Dirac manner produces the following
results. The canonical commutators are
\begin{equation}
[\varphi(\vec x),\Pi(\vec {y})] = i \delta^{2}(\vec {x}-\vec
{y}),\label{dieciocho}
\end{equation}

\begin{eqnarray}
[A_{i}(\vec x),A_{j}(\vec y)]= \frac ik \,
\varepsilon_{ij}\,\delta^{2}(\vec x - \vec y), \label{diecinueve}
\end{eqnarray}
and the Hamiltonian is
\begin{equation}
H= \int d^{2}x \,\frac 12 \,
\Big(\Pi^{2}+(A_{i}+\partial_{i}\varphi)(A_{i}+\partial_{i}\varphi)\Big)
. \label{veinte}
\end{equation}
There is also a first class constraint

\begin{equation}
 k \varepsilon ^{ij}\partial_{i}A_{j}+\,\Pi=0  ,\label{veintiuno}
\end{equation}
that generates the time-independent gauge transformations on the
canonical variables. At this point, it is worth comparing the SDM
with the $2+1 $ dimensional massles scalar field theory , whose
action and Hamiltonian can be obtained by putting $A_{\mu}= 0$ in
eqs. \eref{quince} and \eref{veinte} respectively. Also, the
canonical commutators of the scalar theory are precisely given by
eq.\eref{dieciocho}. Since in this case the gauge symmetry is
absent, there are no constraints (it would be incorrect to set
$A_i=0$ in eq. \eref{veintiuno} and to say that $\Pi=0$ is a
constraint in this case). We shall exploit these apparent
simmilarities by working in a geometric representation based on
the RLs space and employing old ideas borrowed from the loop
representation formulation of gauge theories. Since the theories
we are considering are both Abelian, we need to "abelianize "
 the group of RLs. To this end, we choose the following route.
Given a RL (as in eq. \eref{uno}), we define its "form factor"

\begin{equation}
\rho (\vec x,X)\equiv \sum_{a=1}^{r} s_{a} \,\delta ^{2}(\vec x-
\vec x_a) ,\label{veintidos}
\end{equation}
which allows us to group the RLs accordingly with the following
rule: two RLs are said to be equivalent if they share the same
form factor. It can be easily checked that this indeed defines an
equivalence relation. Moreover, since

\begin{equation}
\rho (\vec x,X.Y)= \rho (\vec x,X)+ \rho (\vec x,Y),
 \label{veintitres}
\end{equation}
each equivalence class of RLs defines an element of an Abelian
group. What we have done is to relax the condition that demanded
points and antipoints to be consecutive (apart from being at the
same place) in order to annihilate each other. In other words,
with this further identification we are not concerned about the
order of the points in the list. Within this  geometric setting
the quantum algebra of the massless scalar field theory can be
realized as follows:

\begin{eqnarray}
exp \,\,(-i\varphi(\vec x)) \Psi(X)&=& exp \,\,(-i\varphi(\vec
x)) \Psi(\vec x_{1}^{\,(s_1)},\vec x_{2}^{\,(s_2)},...\vec x_{r}^{\,(s_r)})\nonumber\\
&=& \Psi(\vec x^{\,(+)},\vec x_{1}^{\,(s_1)},\vec
x_{2}^{\,(s_2)},...\vec x_{r}^{\,(s_r)}), \label{veinticuatro}
\end{eqnarray}

\begin{equation}
\Pi_(\vec {x}) \Psi(X)= \rho (\vec x,X)\Psi(X),
\label{veinticinco}
\end{equation}
as can be verified. Eq.\eref{veinticuatro} amounts to realize

\begin{equation}
\partial_{i}\varphi (\vec x) \rightarrow i\,\delta_{i}(\vec x).
\label{veintiseis}
\end{equation}
From eqs. \eref{veinticuatro}, \eref{veinticinco} we see that in
the space of functionals that depend on Abelian RLs, the operator
$exp \,\,(\mp i\varphi(\vec x))$ appends a "positive" ("negative")
point to the list $X$, while $\Pi (\vec {x})$ displays the form
factor of $X$. It should be observed that it is the derivative of
the field operator (and not the field itself) which enters in the
expressions for the observables of the theory. This is
reminiscent of the invariance of the theory under the shift
$\varphi \rightarrow \varphi + constant$. This derivative is
realized as $i$ times the "dipole" derivative discussed before,
accordingly with eq. \eref{veintiseis}. In terms of $X-$dependent
functionals, the Schr\"{o}dinger equation of the massles scalar
theory becomes

\begin{equation}
i\frac{\partial}{\partial t}\Psi(X,t) = \int d^{2}x \,
\frac{1}{2}\,\,\Big( \rho ^{2}(\vec x,X)- \delta_{i}(\vec
x)\delta_{i}(\vec x)\Big) \Psi(X,t). \label{veintisiete}
\end{equation}
The Hamiltonian comprises a "dipole" Laplacian, together with a
potential term $\rho ^2$ wich should be regularized, since it is
essentially the square of Dirac`s delta functions.

 At this point,
we turn back to the SDM, and try to realize its quantum algebra
in the RLs representation. First of all, notice that we are
dispensed of realizing  gauge-dependent operators. Hence, we
focus on the algebra of the basic gauge-invariant ones
\begin{equation}
[\Pi (\vec x)\, ,\, (A_{i}+\partial_i \varphi) (\vec y)\,]= -i\,
\frac {\partial}{\partial y^{i}} \, \delta^{2}(\vec x - \vec y),
\label{veintiocho}
\end{equation}

\begin{equation}
 [(A_{i}+\partial_i \varphi)(\vec x)\, ,\, (A_{j}+\partial_j \varphi)(\vec y)]= \frac ik \,
\varepsilon_{ij}\,\delta^{2}(\vec x - \vec y).\label{veintinueve}
\end{equation}

It can be seen that the prescriptions
\begin{equation}
\Pi (\vec x)\rightarrow \rho (\vec x,X), \label{treinta}
\end{equation}
\begin{equation}
(A_{i}+\partial_i \varphi)(\vec x) \rightarrow iD_{i}(\vec x)
\equiv i\delta_i (\vec x) + \frac {1}{2\pi k} \sum _{a} s_{a}
\varepsilon_{ij} \frac{(x-x_{a})^{j}}{|\vec x - \vec x_{a}|^{2}},
\label{treintiuno}
\end{equation}
verify eqs. \eref{veintiocho},\eref{veintinueve} when acting on
(Abelian) RLs-dependent wave functionals $\Psi(X)$. It should be
noticed that the second term in the r.h.s. of
eq.\eref{treintiuno} is a genuine RLs-dependent quantity. This is
mandatory in order to have a consistent realization of the
quantum algebra. On the other hand, it must be said that this
term already appears in earlier discussions about anyons in
ordinary quantum mechanics \cite{ocho}.

Using the Abelian version of eq. \eref{trece}

\begin{equation}
\varepsilon^{ij}\partial_{i} \delta_{j}(\vec x) = 0 ,
\label{treintidos}
\end{equation}
it can be shown that the gauge constraint \eref{veintiuno} is
automatically satisfied. It could be interesting to compare this
feature with what occurs in other gauge theories. In the
loop-space formulation of Maxwell Theory \cite{nueve}, it is
obtained that the very introduction of loops (i.e., closed
Faradays lines) suffices to solve the Gauss constraint.
Introduction of point sources, demands that there must be open
Faradays lines, starting or ending at the points where charges
(that must be quantized) are placed \cite{diez}. A similar result
holds when the Proca-Stueckelberg model is quantized in an
appropriate geometric space \cite{once}. In the MCSM, (that is
dual to the SDM that we are considering), however, it was found
that the quantization in path space does not lead to convert the
gauge-constraint in an identity \cite{seis}. Nevertheless, after
solving this constraint in path space it was seen that the
property of paths that really matters is the winding number of
the open curves around their boundaries. Then, performing certain
unitary transformation, it was obtained that this dependence can
be rewritten as a functional dependence in the boundaries of the
paths, together with the inclusion in the Hamiltonian of a term
describing a long-range interaction between these boundaries. But
this is precisely what we have obtained in the present approach,
following a different way. In fact, substituting eqs.
\eref{treinta} and \eref{treintiuno} into the Hamiltonian
\eref{veinte}, we can write the Schr\"{o}dinger equation of the
SDM, in the RLs-representation as
\begin{equation}
i\frac{\partial}{\partial t}\Psi(X,t) = \int d^{2}x \,
\frac{1}{2}\,\,\Big( \rho ^{2}(\vec x,X)- D_{i}(\vec x)D_{i}(\vec
x)\Big) \Psi(X,t), \label{treintitres}
\end{equation}
that differs from \eref{veintisiete} in the appearance of the
covariant derivative $D_{i}(\vec x)$, that encodes the
Chern-Simons interaction. As in the MCSM, one can perform the
singular gauge transformation
\begin{equation}
\Psi(X)\rightarrow \overline{\Psi}(X)\equiv exp\,\,
[i\Lambda(X)]\Psi(X), \label{treinticuatro}
\end{equation}
with
\begin{eqnarray}
\Lambda(X) &=& \frac{1}{4\pi k} \int d^{2}x \int d^{2}y
\,\,\rho(X,\vec x )
\,\,\theta(\vec x - \vec y)\,\,\rho(X,\vec y )\nonumber\\
&=& \frac{1}{4\pi k} \sum_{a} \sum_{a'} s_{a}s_{a'}\,\theta(\vec
x_{a} - \vec x_{a'})\ ,\label{treinticinco}
\end{eqnarray}
to convert the covariant derivative $D_{i}(\vec x)$ into an
ordinary "dipole derivative" $\delta_{i}(\vec x)$, as it appears
in the Schr\"{o}dinger equation of the massless scalar field. In
the last equation, $\theta(\vec x)$ is the angle that $\vec x$
makes with the $x$ axis. The price for this simplification is
that the resulting wave functional $ \overline{\Psi}(X)$ is
multivalued, due precisely to this dependence in the angle [In eq.
\eref{treinticinco} there appear "self-interaction" terms,
proportional to $\theta (\vec 0\,)$, that are not well defined.
We shall ignore these regularization issues in this paper (for
further details see \cite{seisprima})]. Thus, we see that the SDM
can be seen as the theory of a massless scalar field obeying
anyonic boundary conditions, as it happens with the MCST
\cite{seis}.

It should be understood that despite the appearances, there is a
fundamental difference between the two "gauges" that admits the
SDM. In the usual one, the Stueckelberg fields $\varphi$ are
eliminated by means of a legitimate gauge transformation.
Instead, in the second one, the vector field $A_{\mu}$ is
eliminated, but by means of a \emph{singular} gauge
transformation. Nevertheless, there is nothing wrong with this
last point of view, as far as we keep in mind that the wave
functionals become multivalued. To conclude, we want to underline
that this second approach, which is well known for the case of
particles in Chern-Simons interactions \cite{ocho}, becomes quite
natural in the SDM thanks to the introduction of the RLs
formalism.

This work was supported by \emph {Consejo de Desarrollo
Cient\'{\i}fico y Human\'{\i}stico}, Universidad Central de
Venezuela, Caracas, VENEZUELA, and by Project G-2001000712 of
FONACIT, VENEZUELA.

\end{document}